# Quantum interference effects in chemical vapor deposited graphene

Nam-Hee Kim[a], Yun-Sok Shin[a],* Serin Park[b], Hong-Seok Kim[a], Jun Sung Lee[a],

Chi Won Ahn[c], Jeong-O Lee[b], Yong-Joo Doh[a],*

[a]*Department of Applied Physics, Korea University Sejong Campus, Sejong 339-700, Korea*

[b]*NanoBio Fusion Research Center, Korea Research Institute of Chemical Technology, Daejeon 305-343, Korea*

[c]*Nano-Materials Laboratory, National Nanofab Center, Daejeon 305-806, Korea*

**Abstract**

We report several quantum interference effects in graphene grown by chemical vapor deposition. A crossover between weak localization and weak antilocalization effects is observed when varying the gate voltage and we discuss the underlying scattering mechanisms. The characteristic length scale for phase coherence is compared with that estimated from universal conductance fluctuations in the micropore-formed graphene sample. These extensive temperature- and gate-dependent measurements of the intervalley and intravalley scattering lengths provide important and useful insight for the macroscopic applications of graphene-based quantum devices.

Corresponding authors

*E-mail address*: yunsokshin@korea.ac.kr, yjdoh@korea.ac.kr

# 1. Introduction

Graphene[1], a monolayer honeycomb lattice of carbon atoms, provides a unique platform for studying two-dimensional relativistic quantum physics and for developing novel quantum-information devices. The half-integer quantum Hall effect [2, 3] and Klein tunneling [4] were demonstrated in electrical-transport measurements, and graphene-based supercurrent transistors [5, 6] and spintronic devices [7] were realized using mechanically exfoliated graphene. Recent advances in growth techniques of large-scale graphene films [8, 9] by chemical vapor deposition (CVD) open the possibility of macroscopic applications of graphene-based quantum devices for integrated circuits. This provides a strong motivation for investigating the phase-coherent electronic-transport properties of CVD-grown large-scale graphene.

Previous studies of phase-coherent transport in graphene involved the magnetoconductivity (MC) measurement of exfoliated graphene flakes [10-14], which depends on inelastic and elastic scattering of charge carriers in graphene [15, 16]. Since graphene exhibits a chiral nature, where the crystal momentum is coupled to the isospin due to a sublattice degeneracy, the backscattering of charge carriers is reduced. An unusual Berry phase of $\pi$ is expected during coherent backscattering in each valley, resulting in destructive interference [15]. Thus, the weak antilocalization (WAL) in graphene results in negative MC curve [11]. When the sublattice degeneracy in graphene is broken by atomically sharp scatterers (e.g., sample edges or ridges), elastic intervalley scattering restores weak localization (WL), as evidenced by a positive MC [12, 13]. Another type of elastic scattering is intravalley scattering due to ripples [16] and trigonal warping [15]; here, isospin is not conserved. Then, the effective time-reversal symmetry (TRS) is broken in each valley and WAL is suppressed [10, 14].

In comparison with mechanically exfoliated graphene, there have been few MC measurements on CVD-grown graphene by varying temperature [17], or gate voltage [18], or strain [19], respectively. We here report an extensive study of phase-coherent electronic transport in CVD-grown and transferred large-scale graphene that shows clear gate-voltage and temperature dependences. The overall WL behavior is converted to the WAL feature near the charge-neutrality point (CNP) and the phase coherence length is compared with the one estimated from universal conductance fluctuations (UCFs) in graphene. Our observations provide important insight into quantum-electronic transport in CVD-grown and transferred graphene that is ultimately relevant for the development of novel quantum-information devices based on large-scale graphene.

## 2. Experiments

The growth and device-fabrication processes for graphene samples are illustrated in Fig. 1a-e. A Cu foil (10 × 10 mm$^2$) is mounted at the center of a quartz tube in a hot wall furnace under vacuum. After annealing the Cu foil at 1,000 ℃, graphene growth is initiated with a mixture of $H_2$ and $CH_4$ gases flowing through the quartz tube. After a continuous large-scale graphene film is formed on the Cu foil, the furnace is cooled down to room temperature (Fig. 1a). The graphene film is coated with a thin layer of polymethyl methacrylate (PMMA) and then it is baked at 120 ℃ to evaporate the solvent. The Cu foil is removed using a Cu etchant, leaving only the PMMA/graphene film (Fig. 1b). After the film is cleaned in a bath of deionized water, it is transferred onto a $SiO_2$/Si substrate (Fig. 1c). The PMMA film is removed with acetone, leaving the graphene layer on the $SiO_2$/Si substrate (Fig. 1d). A patterned film of Al is formed onto the graphene layerby photolithography and thermal evaporation of a 30-nm-thick Al film. The resultant pattern of Al/graphene film is formed by $O_2$ plasma etching of the graphene regions not masked by the Al film. After etching the Al film (etchant: AZ300 MIF), the patterned graphene is achieved (Fig. 1e). Three-terminal graphene devices contacted with Cr/Au (3 nm/15 nm) electrodes are fabricated by electron-beam lithography (Fig. 1f). Electrical transport properties are characterized in a two-point measurement configuration with and without a magnetic field ($B$), while a source-drain bias ($V_{sd}$) and a back-gate voltage ($V_g$) are applied.

After completion of the device fabrication, an atomic force microscopy (AFM) image and micro-Raman spectroscopy (532-nm laser excitation) results are compared directly to confirm the uniform thickness of the CVD-grown graphene film. The spatial map of the intensity ratio between the 2D (~ 2,685 cm$^{-1}$) and the G (~ 1,584 cm$^{-1}$) band Raman peaks, $I_{2D}/I_G$, gives values in excess of 2.0 over the entire graphene film in Fig. 2b, indicating uniform monolayer graphene [20]. Representative Raman spectra, obtained at three different locations on the film, are displayed in Fig. 2c. We note that the peak heights of the G band are almost the same, whereas those of the 2D band vary widely. The relatively large variation of $I_{2D}$ is attributed to the spatially non-uniform adhesion between the transferred graphene and the substrate [21]. We also note that the disorder-induced D band (~ 1,346 cm$^{-1}$) peak is very low or absent, reflecting the high quality of the CVD-grown graphene film [20].

3. **Results and discussion**

The temperature dependence of the sheet resistance $R_{sheet}$ of the graphene sample is plotted in Fig. 2d as a function of $V_g$. It reveals that the charge neutrality point (or Dirac point) $V_{CNP}$, at which the electron and hole concentrations are equal, shifts from $V_g =$ 34.2 V at room temperature to $V_g$ = 27.6 V at $T$ = 2.6 K. The carrier mobility and the mean free path at $T$ = 2.6 K are estimated as $\mu$ = 1200 (2000) cm$^2$/Vs and $l_m$ = 18 (27) nm, respectively, for $\Delta V_g$ = -20 (20) V, where $\Delta V_g = V_g - V_{CNP}$. These values are very similar to those obtained from a mechanically exfoliated graphene film [22]. The carrier concentration can be estimated from the relation [2] $n = 7.2\times10^{10}|\Delta V_g|$ (cm$^{-2}$), which yields $n = 2\times10^{12}$ cm$^{-2}$ for $|\Delta V_g|$ = 20 V. In a strong magnetic field of $B$ = 9 T, our graphene device exhibits conductance plateaus at $G = \nu e^2/h$ with $\nu$ = 2 and 6, where $e$ is the elementary charge and $h$ is the Planck constant, as shown in the inset of Fig. 2d. Since the longitudinal conductivity is mixed with the Hall conductivity in a two-point measurement configuration [23], those plateaus are attributed to a "half-integer" quantum Hall effect [2, 3], which is again indicative of single-layered graphene.

Fig. 3a shows the differential MC, $\Delta\sigma = \sigma(B) - \sigma(B=0)$, at different temperatures under a fixed $V_g$ = 0. We note that CVD-grown graphene displays positive MC, a typical feature of WL, and the positive MC correction clearly increases at lower temperatures. Quantitative analysis of the WL-induced conductivity correction is made possible by the theoretical formula derived by McCann *et al.* [15],

$$\Delta\sigma = \frac{e^2}{\pi h}\left[F\left(\frac{8\pi B}{\Phi_0 L_\phi^{-2}}\right) - F\left(\frac{8\pi B}{\Phi_0 \{L_\phi^{-2} + 2L_i^{-2}\}}\right) - 2F\left(\frac{8\pi B}{\Phi_0 \{L_\phi^{-2} + L_i^{-2} + L_*^{-2}\}}\right)\right]. \quad (1)$$

Here $F(z) = \ln(z) + \psi(0.5+z^{-1})$, $\psi(x)$ is the digamma function, and $\Phi_0 = h/e$ is the magnetic-flux quantum. The characteristic length scales $L_\phi$ and $L_i$ represent the phase-coherence length (or inelastic scattering length) and the elastic intervalley scattering length, respectively, while $L_*$ means the elastic intravalley scattering length. The latter is given by $L_* = \sqrt{D\tau_*}$, where $D$ is the diffusion constant and $\tau_*$ is the intravalley scattering time. Since $D = v_F l_m/2$, where $v_F$ is the Fermi velocity, we obtain $D = 90$ cm$^2$/s at $V_g = 0$ V. Notably, the first term in Eq. (1) is responsible for the positive MC (or WL), while the other terms produce negative MC (or WAL).

Fitting the experimental MC data to the McCann formula (Fig. 3a) returns the scattering lengths $L_\phi$, $L_i$ and $L_*$ of the graphene sheet as functions of temperature. At $T = 2.6$ K, they are estimated to be $L_\phi = 320$ nm, $L_i = 430$ nm, and $L_* = 17$ nm, as shown in Fig. 3b. These values relate to the inelastic scattering time $\tau_\phi = 11$ ps, the intervalley elastic scattering time $\tau_i = 21$ ps, and the intravalley elastic scattering time $\tau_* = 32$ fs, respectively, via $L_{\phi,i,*} = \sqrt{D\tau_{\phi,i,*}}$. These values are similar to earlier results obtained from mechanically exfoliated [13] and CVD-grown [19] graphene films. It is clear that $L_\phi$ decreases with increasing temperature, while $L_i$ and $L_*$ are almost temperature-independent in our experimental range. Since $L_\phi$ is proportional to $T^{-0.5}$ (Fig. 3b), the electron dephasing rate $\tau_\phi^{-1}$ should exhibit a positive linear dependence on temperature. Since electron-phonon scattering is expected to be weak in graphene [14], the electron-electron interaction, the so-called Nyquist scattering, is considered to be the major cause of inelastic scattering [12, 13], resulting in an electron dephasing rate of $\tau_\phi^{-1} = \alpha k_B T \ln g/\hbar g$, where $\alpha$ is an empirical coefficient between 1 and 2, $k_B$ is the Boltzmann constant, and $g = \sigma h/e^2$ is the normalized conductivity. Our experiment places $\alpha$ at 1.0 for $T = 2.6$ K and $g(V_g = 0) = 7.0$, indicating that the Nyquist scattering would be a major cause of inelastic scattering in CVD-grown graphene. Another inelastic scattering mechanism, such as the direct Coulomb interaction between electrons [14], which produces a parabolic dependence on temperature, seems not to play a dominant role in the overdoped regime at $V_g = 0$ V.

The evolution of the measured MC with varying $V_g$ is displayed in Fig. 3c. When $V_g$ approaches the CNP, $V_{CNP} = 27.6$ V, the positive MC behavior becomes suppressed. At magnetic fields greater than $B \sim 0.1$ T, the MC exhibits a downturn at $V_g = 26$ V. This negative MC behavior is attributed to WAL in graphene [13, 14, 24]. An analysis of the gate-dependent $\Delta\sigma$ based on Eq. (1) reveals the gate-voltage dependence of $L_\phi$, $L_i$, and $L_*$. Fig. 3d shows that $L_\phi$ and $L_i$ decrease with the carrier density, while $L_*$ increases near the CNP. In terms of the scattering time, $\tau_i$ ($\tau_\phi$) changes from 20 to 10 (11 to 3) ps and $\tau_*$ from 0.04 to 0.2 ps, respectively, as $V_g$ approaches the CNP from $V_g = 0$. We also note that $L_i$ is always greater than $L_\phi$ in the entire $V_g$ range, signifying that the intervalley scattering rate is less than the phase-breaking rate ($\tau_i^{-1} < \tau_\phi^{-1}$). The reduced intervalley scattering rate, also observed in other CVD-grown samples [18, 19], can be caused by the graphene being loosely attached to the substrate. This behavior is in striking contrast to the mechanically exfoliated graphene, which is tightly coupled to the Si substrate and thus exhibits much stronger elastic intervalley scattering [12, 15].

The decrease in $L_\phi$ with decreasing carrier density $n$ is consistent with Nyquist-type electron-electron scattering [12], resulting in $L_\phi$ being proportional to $(g/\ln(g))^{0.5}$. The theoretical value $L_\phi$ = 270 nm expected from the dephasing rate at the CNP, however, is much larger than the value $L_\phi$ = 165 nm estimated from the WAL fitting. This discrepancy suggests an additional dephasing mechanism that becomes activated near the CNP, e.g., a direct Coulomb interaction due to poor screening [12] or electron-hole puddles formed at the CNP [12], which reduce the effective dimension of the sample [13]. The decrease in $L_i$ near the CNP can be understood in terms of weak screening near the CNP [15], which enhances intervalley scattering [18].

The occurrence of negative MC, as an indicator of WAL, is directly related to the smallness of the ratios $\tau_\phi/\tau_*$ and $\tau_\phi/\tau_i$ [14], which can be deduced from Eq. (1). Consequently, our observation of WAL near the CNP is accompanied by a drastic decrease in $\tau_\phi/\tau_*$, which varies from 240 at $V_g = 0$ to 14 at the CNP, while $\tau_\phi/\tau_i$ diminishes moderately from 0.55 to 0.33. The drastic change in $\tau_\phi/\tau_*$ is mainly attributed to an abrupt increase of $\tau_*$ (and correspondingly $L_*$) near the CNP in addition to the monotonous decrease of $\tau_\phi$ while approaching the CNP. This gate dependence of $\tau_*$ can be useful for quantifying the intravalley scattering mechanism in CVD-grown graphene. Since $L_*$ is much smaller than $L_i$, the atomically sharp scatterers in a graphene sheet, which can be a common cause of the intra- and inter-valley scatterings, cannot explain the observed large intravalley scattering rate in our experimental range [13]. There must therefore be another scattering process that affects $L_*$ but not $L_i$.

Corrugation-like features ("ripples") in natural graphene [25] can also cause intravalley scattering by inducing pseudo-magnetic fields and effectively breaking time-reversal symmetry within a valley in graphene [16]. This ripple-induced scattering rate is given by $\tau_*^{-1} \sim v_F/k_F d^2$, where $k_F$ is the Fermi wave vector and $d$ the ripple diameter [16]. Because $\tau_*^{-1}$ diverges at the CNP [6], ripple-induced scattering is inconsistent with our experimental observations in CVD-grown graphene. An intravalley scattering rate induced by randomly distributed dislocations would be equally inconsistent [13]. Another possible cause of intravalley scattering would be the trigonal warping caused by the anisotropy of the Fermi surface in $k$ space. Since the trigonal warping scattering rate $\tau_w^{-1}$ is proportional to the square of the carrier density [15], the corresponding scattering length $L_*$ should decrease linearly with the carrier density or with $\Delta V_g = |V_g - V_{CNP}|$ near the CNP [17], consistent with our observations. The theoretical value of $\tau_w$, however, is found to be 450 ps at the CNP [18], which is three orders of magnitude larger than that expected by MC measurements in this work. We therefore infer that trigonal warping alone cannot explain the strong intravalley scattering in CVD-grown graphene.

Local potential fluctuations induced by charged impurities in graphene can also induce intravalley scattering, strongly depending on the carrier density [13]. Charged impurities could include $H_2O$ absorbed at the graphene/substrate interface or the ionic residue left over from the Cu-foil etching process [18]. Inhomogeneous electron-hole puddles formed near the CNP can induce similar potential fluctuations, which break the effective time-reversal symmetry in each valley. It is known that the potential-induced scattering rate is proportional to the number of carriers per unit cell of graphene [16]. When the carrier density decreases near the CNP, the Coulomb interaction between the charged impurities and the carriers is reduced and hence $\tau_*$ increases, consistent with our observations. However, a quantitative estimate [13] of $\tau_*$ is approximately two orders of magnitude greater than the estimates based on our experimental MC curves. Thus, a more detailed theory is required to understand the mechanism of strong intravalley scattering in CVD-grown graphene.

Low-temperature MC curves, obtained from a different sample containing three micropores of diameter 1.2 μm in the middle of the graphene layer, are plotted in Fig. 4a for different angles $\theta$ between the magnetic field and the substrate plane. When the different MC curves are replotted as functions of the perpendicular magnetic field, $B\sin\theta$, they map onto a single curve that is attributed to the two-dimensional nature of graphene. We note that quasi-periodic conductance oscillations are superposed onto the background MC curve, which were not observed in the samples without the holes. We subtracted the MC background to isolate the conductance oscillations $\delta G$ for the different $\theta$. Resultantly, the average periodicity turns out to be $\Delta B = 0.18 \pm 0.033$ T.

One possible explanation for the conductance oscillations would be the Aharonov-Bohm (AB) effect [26], typically observed in mesoscopic rings. The phase interference between two electron partial waves propagating on different sides of the ring is modulated by a perpendicular magnetic flux threading the interior of the ring with a periodicity given by the magnetic flux quantum $\Phi_0 = h/e$. Assuming this scenario, we would expect a periodicity $\Delta B = 3.7$ mT, based on the hole area in the graphene layer. However, this is almost two orders of magnitude smaller than our observation. The AB effect is therefore ruled out.

A more plausible explanation for the superposed conductance oscillations is UCFs in the narrow channels between the micropores in the graphene sample. UCFs are caused by the quantum interference of multiply scattered electronic wavefunctions in a weakly disordered conductor, giving rise to reproducible and aperiodic conductance fluctuations with an amplitude of order $e^2/h$ as a function of the magnetic field [27]. Figure 4b shows the autocorrelation function of $\delta G$, defined as $F(\Delta B) = \langle\delta G(B)\delta G(B+\Delta B)\rangle - \langle\delta G(B)\rangle^2$, with $\Delta B$ being a lag parameter in the magnetic field. The half width at half maximum of $F(\Delta B)$ corresponds to the magnetic correlation length, $B_c$, over which the quantum interference becomes incoherent [28]. Figure 4b suggests that $B_c = 0.021$ T. Since $B_c = 0.95\Phi_0/L_\phi^2$ for a two-dimensional conductor [29], where the phase coherence length is estimated to be $L_\phi = 430$ nm, comparable to $L_\phi = 240$ nm estimated by fitting the MC curve obtained from the microporous graphene to the McCann formula (see the inset). Thus our observations of UCFs and of WL/WAL provide direct evidence for coherent quantum electron transport in CVD-grown graphene.

4. **Conclusion**

In summary, quantum interference effects in CVD-grown and transferred graphenewere investigated extensively by varying the temperature or gate voltage. A thorough analysis of MC data reveals that WL in overdoped region converts into WAL near the CNP, which is caused by enhanced intervalley scattering and reduced intravalley scattering near the CNP. $L_i$ and $L_*$ are quite insensitive to temperature, while $L_\phi$ decreases monotonously with temperature as a result of the electron-electron interaction. Measurements of UCFs confirms the phase coherent transport in the CVD-grown graphene. Our observations would be a promising feature for the large-scale integration of graphene-based quantum information devices in the near future.

**Acknowledgment**

This work was supported by the National Research Foundation of Korea through the Basic Science Research Program (Grant No. 2012R1A1A2044224 for YSS and 2015R1A2A2A01006833 for YJD).

**Figure captions**

**Fig. 1.** Synthesis, transfer, and device fabrication processes for large-scale graphene film. (a) CVD growth of graphene film on a Cu foil (~ 10×10 mm$^2$). (b) Separation of a polymethyl methacrylate (PMMA)/graphene film from a Cu foil using a Cu etchant. (c) Transfer of PMMA/graphene film onto a SiO$_2$/Si substrate. (d) Removal of PMMA using acetone. (e) Micropatterning of a graphene film by oxygen plasma etching using an Al mask. The mask is removed with an Al etchant. (f) Deposition of metallic electrodes by electron-beam lithography and electron-beam evaporation. The source and drain electrodes are made of a Cr/Au double layer, while the highly doped Si substrate is used as a back gate. A magnetic field $B$ is applied at various inclinations relative to the substrate.

**Fig. 2.** (a) AFM image of CVD-grown graphene sample **S1**. The edge of a Cr/Au electrode is visible at the top of the image. (b) Spatial map of the Raman intensity ratio of $I_{2D}/I_G$. Each Raman spectrum was taken in 1-μm increments in the *x* and *y* directions. (c) Representative Raman spectra obtained at different locations on the graphene film, indicated by the circle, triangle, and square symbols in (b). (d) Temperature dependence of the sheet resistance $R_{sheet}$ as a function of the gate voltage $V_g$. Inset: $V_g$-dependent

two-terminal conductance in a graphene sample **S2** with $B = 9$ T at $T = 2.6$ K. The conductance plateaus at $G = \nu e^2/h$ ($\nu = 2, 6$) are due to the half-integer quantum Hall effect in graphene (see text).

**Fig. 3.** (a) Differential magnetoconductivity, $\Delta\sigma = \sigma(B) - \sigma(0)$, of sample **S1** at temperatures $T = 5, 10, 15, 40$, and $70$ K (top to bottom). The solid curves show the best fits for $\Delta\sigma$ using the McCann formula (see text). (b) The characteristic lengths $L_\phi$, $L_i$, and $L_*$, as functions of $T$. $L_\phi$ obeys the power law $T^{-0.5}$. (c) Differential magnetoconductivity $\Delta\sigma$ for different gate voltages $V_g = 0, 20, 23$, and $26$ V (top to bottom) at $T = 2.6$ K. The solid curves show the best fits for $\Delta\sigma$ based on the McCann formula (see text). (d) The characteristic lengths $L_i$, $L_\phi$, and $L_*$ as functions of $V_g$. Solid lines are drawn only to guide the eye.

**Fig. 4.** (a) Differential magnetoconductance $\Delta G = G(B) - G(0)$ as a function of the perpendicular magnetic field $B\sin\theta$, which was obtained from sample **S3** for different angles $\theta = 90°$ (black), $60°$ (red), $45°$ (green), $30°$ (blue), and $5°$ (cyan). Inset: AFM image of the sample with the micropores. The white dotted lines highlight the hole perimeters for clarity and the scale bar indicates 1 μm. (b) Autocorrelation function

(solid line) obtained from δ$G$ for $\theta$=90° in (a). The dotted line indicates the magnetic correlation length $B_c$. Inset: Δ$\sigma$ (symbol) for $\theta$ = 90°. The solid line is a fit to the McCann formula (see text).

**References**

_

**Fig. 1**

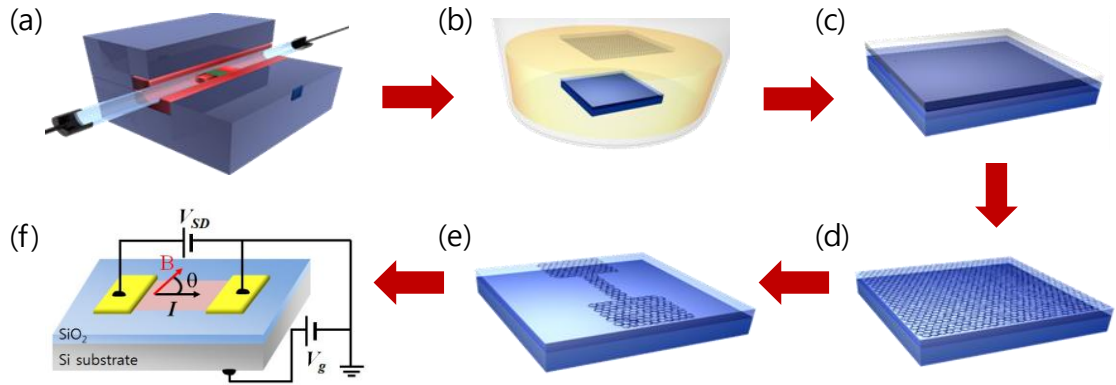

**Fig. 2**

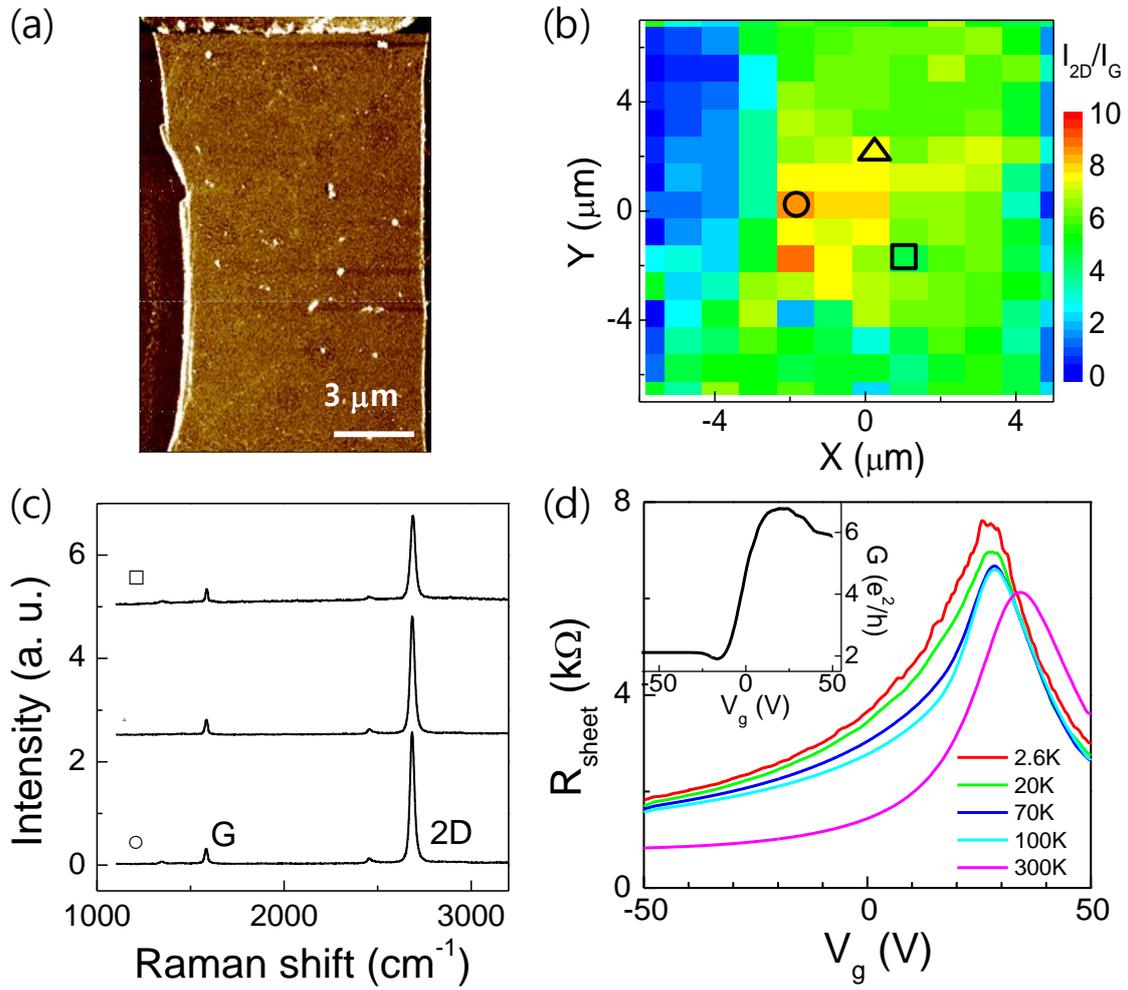

**Fig. 3**

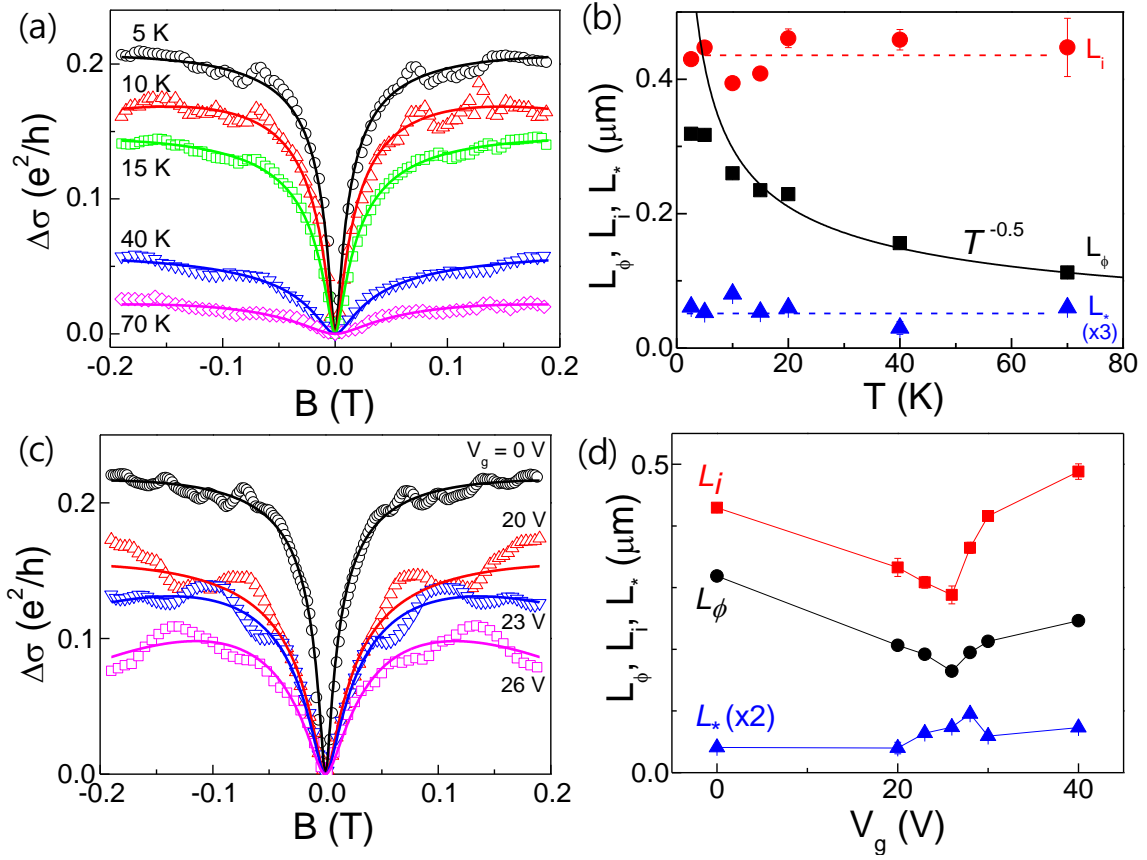

**Fig. 4**

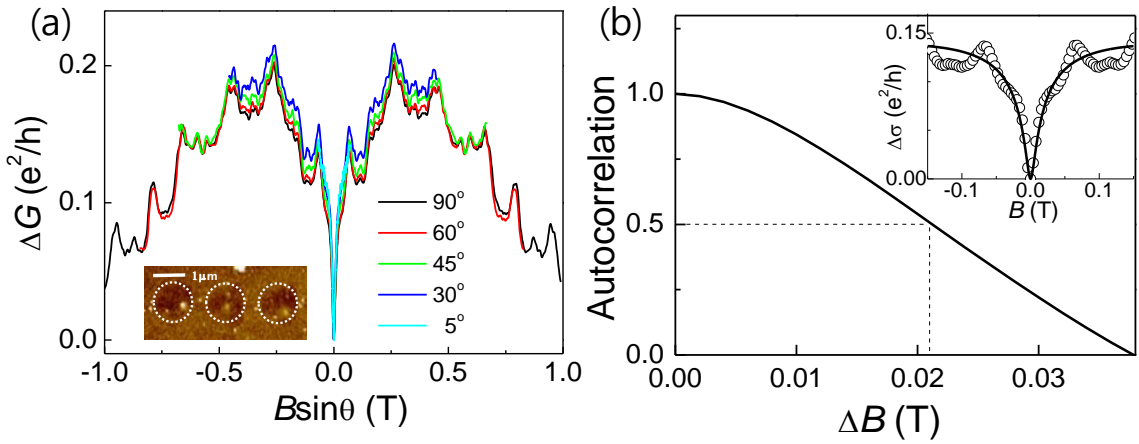